\documentclass[twocolumn,prl,amsmath,letterpaper,floatfix,showpacs]{revtex4}
\usepackage{graphicx}
\usepackage{amsmath}
\usepackage{amssymb}
\usepackage{color}
\usepackage{gensymb}

\newcommand{\figref}[2]{Fig.~\ref{#1}~{\bf #2}}
\newcommand{\figrefs}[1]{Fig.~\ref{#1}}
\newcommand{\sot}[0]{$\mathrm{SO_2}$}

\begin{document}

    \title{Adiabatic field-free alignment of asymmetric top molecules with an optical centrifuge}

    \author{A. Korobenko}
    \email{korobenk@phas.ubc.ca}
    \author{V. Milner}
    \date{\today}

    \affiliation{Department of  Physics \& Astronomy, The University of British Columbia, Vancouver, Canada}

    \begin{abstract}
        We use an optical centrifuge to align asymmetric top \sot{} molecules by adiabatically spinning their most polarizable O-O axis. The effective centrifugal potential in the rotating frame confines sulfur atoms to the plane of the laser-induced rotation, leading to the planar molecular alignment which persists after the molecules are released from the centrifuge. Periodic appearance of the full three-dimensional alignment, typically observed only with linear and symmetric top molecules, is also detected. Together with strong in-plane centrifugal forces, which bend the molecules by up to 10 degrees, permanent field-free alignment offers new ways of controlling molecules with laser light.
    \end{abstract}

    \maketitle

Aligning the axes of gas-phase molecules in the laboratory frame has long been recognized as a powerful instrument in the growing number of areas of molecular science. For recent reviews on the impact of molecular alignment on the molecular dynamics, interactions and spectroscopy, see Refs.\citenum{Ohshima2010, Fleischer2012, Lemeshko2013}. Today, linear and symmetric top molecules are routinely aligned in one dimension with a single laser pulse, either long (adiabatic) or short (non-adiabatic) on the time scale of the rotational period, interacting with the induced dipole moment of a molecule\cite{Stapelfeldt2003a, Seideman2005, Kumarappan2007}.
Aligning the frame of asymmetric top molecules in three dimensions requires more sophisticated methods, capable of controlling the rotation of a molecule around all three of its distinct rotational axes (for recent reviews, see\cite{Artamonov2008, Artamonov2012}). Both adiabatic\cite{Larsen2000, Tanji2005} and non-adiabatic\cite{Underwood2005, Lee2006, Rouzee2008, Pabst2010, Ren2014} excitation schemes, as well as their combinations\cite{Viftrup2007, Viftrup2009}, have been used for the three-dimensional alignment (3DA) of asymmetric rotors.

Adiabatic approaches excel in producing permanent molecular alignment, but only in the presence of a strong laser field, which is often undesirable. Field-free 3DA (FF3DA) have been achieved by means of the non-adiabatic excitation with a single elliptically polarized pulse\cite{Rouzee2008} or a series of pulses\cite{Lee2006, Ren2014}. However, unlike the case of symmetric molecules, where the aligned state is created promptly after the laser pulse and then periodically revives in time, the dynamics of asymmetric rotors exhibit only partial revivals with incommensurate frequencies\cite{Poulsen2004, Holmegaard2007, Spector2014}. As a result, the degree of the prompt post-pulse FF3DA is not only limited\cite{Rouzee2008}, but is also very sensitive to the field ellipticity and strength\cite{Holmegaard2007}. Owing to the same complex aperiodic dynamics, enhancing 3DA with consecutive rotational kicks is inefficient for light molecules like \sot{} \cite{Pabst2010}, and even for heavier rotors, establishing the optimal timing and ellipticity of pulses is not straightforward and may require experiments with feedback-loop based optimization algorithms\cite{Ren2014}.
\begin{figure}[b]
    \includegraphics[width=1\columnwidth]{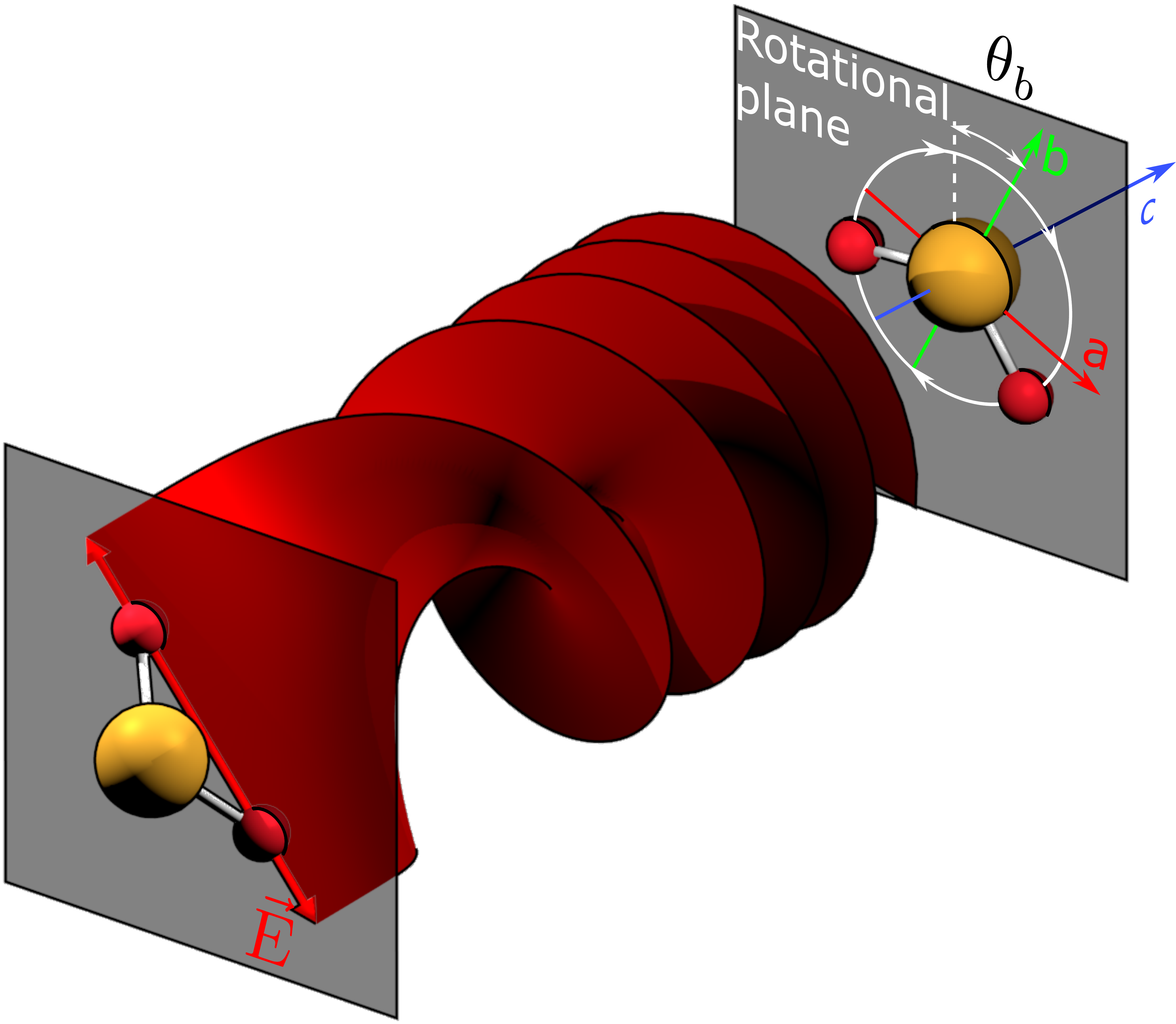}
    \caption{(color online) Illustration of the main concept of aligning an asymmetric top molecule (\sot{}) with an optical centrifuge. Left side of the centrifuge pulse (red corkscrew surface) corresponds to its leading edge, linearly polarized along $\vec{E}$. Behind the trailing edge of the centrifuge (right side), the molecular plane is aligned in the plane of the induced rotation. The three axes of \sot{} ($a,b$ and $c$) are shown in red, green and blue, respectively. $\theta _{b}$ is the orientation angle measured in this work, as discussed in text.}
    \label{fig_so2}
\end{figure}

In this work, we demonstrate a new approach to aligning asymmetric top molecules with non-resonant laser pulses. Our method combines the robustness of an adiabatic alignment with field-free conditions of the final aligned state. It is based on the accelerated spinning of molecules with an optical centrifuge - an intense laser pulse, whose linear polarization rotates with gradually increasing angular frequency\cite{Karczmarek1999, Villeneuve2000}. Optical centrifuge has recently been used for reaching extreme rotational states, known as molecular ``superrotors''\cite{Yuan2011, Korobenko2014a}. High degree of planar alignment has been demonstrated in the ensembles of centrifuged linear molecules\cite{Korobenko2015a}. Here, we extend this technique to the adiabatic alignment of asymmetric rotors.

The essence of our method is illustrated in \figrefs{fig_so2}. The leading edge of the centrifuge (left plane) adiabatically aligns the molecular axis of maximum polarizability ($a$ axis of \sot{}, parallel to the O-O bond) along the field direction. As the rotational frequency $\omega$ increases, slowly emerging centrifugal forces in the rotating frame of reference pull the sulfur atom into the plane of rotation, so as to minimize the effective potential $V_\text{eff}=-\frac{1}{2}I\omega^2$ by maximizing the molecular moment of inertia $I$ around the rotation axis.

\begin{figure}
    \includegraphics[width=\columnwidth]{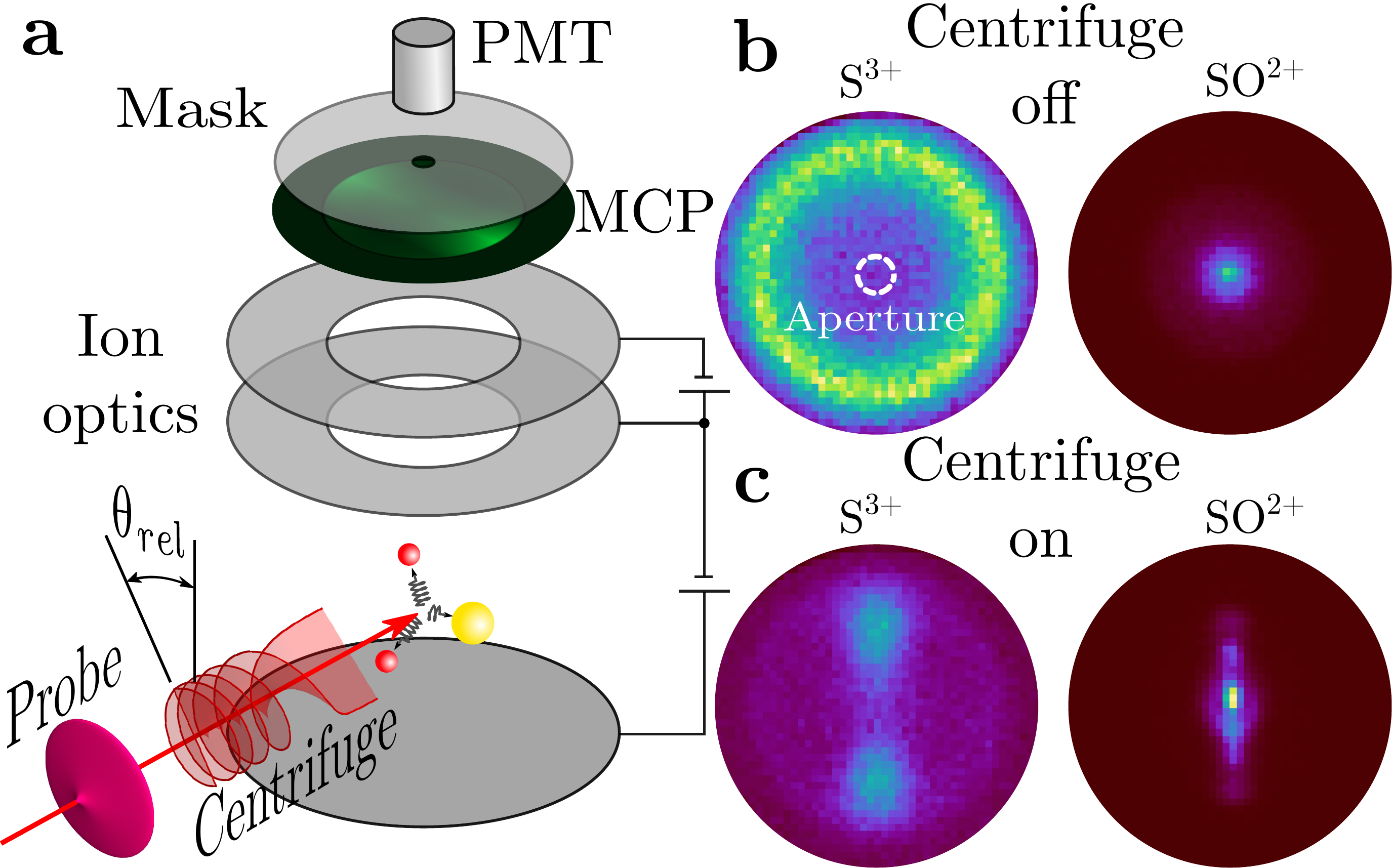}
    \caption{(color online){\bf (a)} Experimental configuration. Supersonically expanded cold molecular ensemble of \sot{} is rotationally excited by an optical centrifuge and Coulomb-exploded by a femtosecond probe beam between the charged plates of a time-of-flight spectrometer. Ion fragments are collected on a microchannel plate detector (MCP) with a phosphorus screen. To determine the in-plane angular distribution of molecules, an opaque mask with a pinhole in the center is placed on top of the screen, with a photomultiplier tube detector (PMT) behind it (see text for details).  {\bf (b,c)} Images of $S^{3+}$ and $SO^{2+}$ fragments originated from the rotationally cold and centrifuged molecules, respectively.}
    \label{fig_setup}
\end{figure}

To observe the field-free rotational dynamics of the centrifuged \sot{} molecules, we employed the technique of velocity map imaging (VMI) in the similar fashion to that described in our previous work\cite{Korobenko2014}. Briefly, a supersonic jet of helium-seeded sulfur dioxide is exposed to the field of an optical centrifuge between the plates of a time-of-flight (TOF) mass spectrometer (\figref{fig_setup}{a}). Rotationally excited molecules are then Coulomb-exploded by an ultrashort intense probe pulse (40~fs, $\sim10^{15}\mathrm{W/cm^2}$) of either linear or circular polarization. The momenta of the fragment atomic and molecular ions, bearing the information on the molecular orientation at the moment of explosion, are projected onto the microchannel plate (MCP) detector with a phosphorus screen.

Gating the MCP voltage for the arrival of either $S^{3+}$ or $SO^{2+}$ ions, we recorded the corresponding VMI images obtained with linear probe polarization, normal to the plane of the MCP. Due to the symmetry of \sot{}, the recoil momenta of both species lie in the O-S-O plane of the fragmented molecule and, therefore, define unambiguously the molecular plane at the time of explosion. Without the centrifuge,  the observed images exhibited an axial symmetry, expected for the spherically symmetric molecular distribution (\figref{fig_setup}{b}). As we turned the centrifuge on, both images collapsed to the plane of the laser-induced rotation, as shown in \figref{fig_setup}{c}. The simultaneous collapse of the two distributions for both ion fragments indicates strong planar alignment of \sot{} molecules.

We studied the rotational dynamics of aligned molecules within the rotation plane using circularly polarized probe pulses to ensure an isotropic in-plane ionization probability. The in-plane angular distribution of centrifuged molecules was measured as a function of the delay time between the end of the centrifuge pulse and the probe-induced Coulomb explosion. Similarly to our previous work\cite{Korobenko2014}, the distribution was determined by recording the ion signal at a single location on the MCP detector (here, in the center of the phosphorous screen) as a function of the release angle of the centrifuge, $\theta _\text{rel}$ (see \figref{fig_setup}{a}).
The latter was found in a separate cross-correlation setup, described in detail in Ref.\citenum{Korobenko2014}.

With the MCP voltage gated at the arrival of $S^{3+}$ ions, the detected signal is proportional to the amount of molecules, $f_{\theta _\text{rel}}(\theta_b \equiv 0)$, with zero orientation angle $\theta_b$ between the reference TOF axis and the molecular $b$ axis (the one bisecting the O-S-O bond angle, \figrefs{fig_so2}). The dependence of this signal on the release angle $\theta _\text{rel}$ is equivalent to the dependence on $\theta_b$ at a fixed $\theta_\text{rel} \equiv 0$, i.e. $f_{\theta_\text{rel}}(0) \equiv f_0(-\theta_b)$. In what follows, we investigate the latter distribution [hereafter referred to as $f(\theta_b)$] as a function of the centrifuge-to-probe delay time.

\begin{figure}
    \includegraphics[width=\columnwidth]{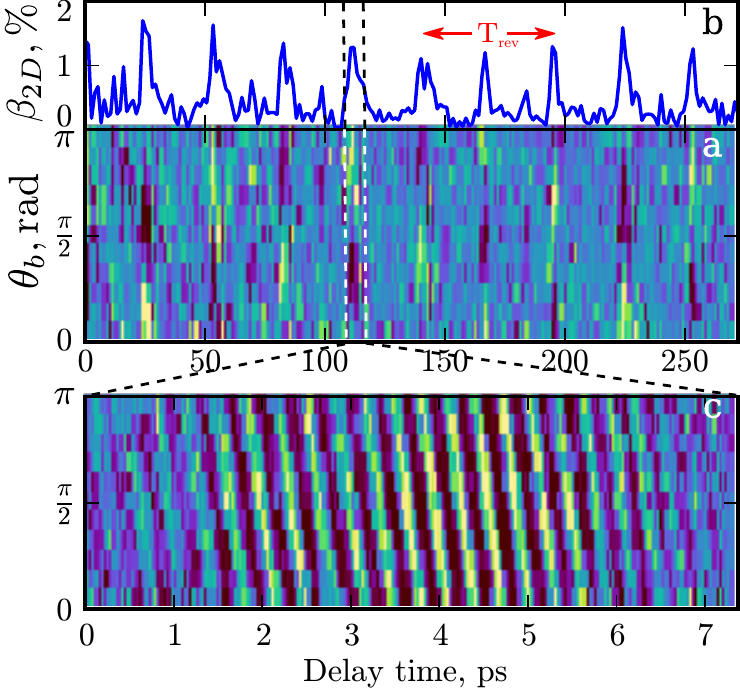}
    \caption{(color online){\bf (a)} Time evolution of the molecular in-plane angular distribution.  {\bf (b)} Periodic revivals of the calculated two-dimensional alignment factor $\beta_{2D}=\langle\cos^2\theta\rangle_{2D}-\frac{1}{2}$. {\bf (c)} High resolution time scan around the alignment peak.}
    \label{fig_rotation}
\end{figure}

The observed time dependence of $f(\theta_b)$ is shown \figref{fig_rotation}{a}. Truncating the centrifuge pulse in time enables us to control the final angular frequency of the centrifuged molecules, which in this case was set to $10^{13}\text{ rad/s}$. Due to the centrifuge-induced planar alignment, the number of the rotational degrees of freedom of \sot{} molecules is reduced to one and their dynamics becomes periodic, similar to the dynamics of a linear rotor. This periodicity, though not apparent from the coarse two-dimensional scan, is clearly evident from the extracted alignment factor, $\beta_{2D}=\langle\cos^2\theta\rangle_\text{2D}-\frac{1}{2}$ (with $\langle..\rangle_\text{2D}$ being the in-plane average), plotted in \figref{fig_rotation}{b}. Each peak corresponds to the field-free three-dimensional alignment of \sot{}.

The relatively small degree of the observed FF3DA stems from the adiabatic mechanism of the centrifuge spinning, which results in a low number of quantum rotational states in the excited wave packet. As we have shown in a recent work on linear rotors\cite{Korobenko2014}, a small number of participating states gives rise to long windows of classical-like rotation around the time of each revival, with the rotational frequency equal to the final frequency of the centrifuge. The same ``stopwatch'' rotation was observed here with \sot{} molecules when we performed a fine time scan around any of the alignment peaks, as demonstrated in \figref{fig_rotation}{c}.

As expected for the effectively one-dimensional rotation, the alignment peaks are separated by half the revival period $T_{rev}=2\pi \hbar (d^2E/dJ^2)^{-1}$, where $E(J)$ is the rotational energy for a given rotational quantum number $J$ and $\hbar$ is the reduced Plank's constant \cite{Averbukh1989,Seideman1999}. Using the rigid rotor's energy spectrum $E(J)=\frac{\hbar^2}{2I}J(J+1)$, with $I$ being the molecular moment of inertia, one finds $T_\text{rev} = 2\pi I/\hbar$. At lower centrifuge frequencies, the experimentally detected revival period of 57~ps corresponds to $I=58~\mathrm{\AA^2\cdot amu}$ in good agreement with the known value for the molecule's largest moment of inertia ($I_{c}$) around the axis normal to its plane ($c$-axis) \cite{Herzberg1966}. These revivals, known as C-type transients\cite{Joireman1992} can be attributed to the beating between a few quantum states $\left\vert J, \tau=-J\right\rangle$, corresponding to the lowest energy for a given $J$.
\begin{figure}[t]
    \includegraphics[width=\columnwidth]{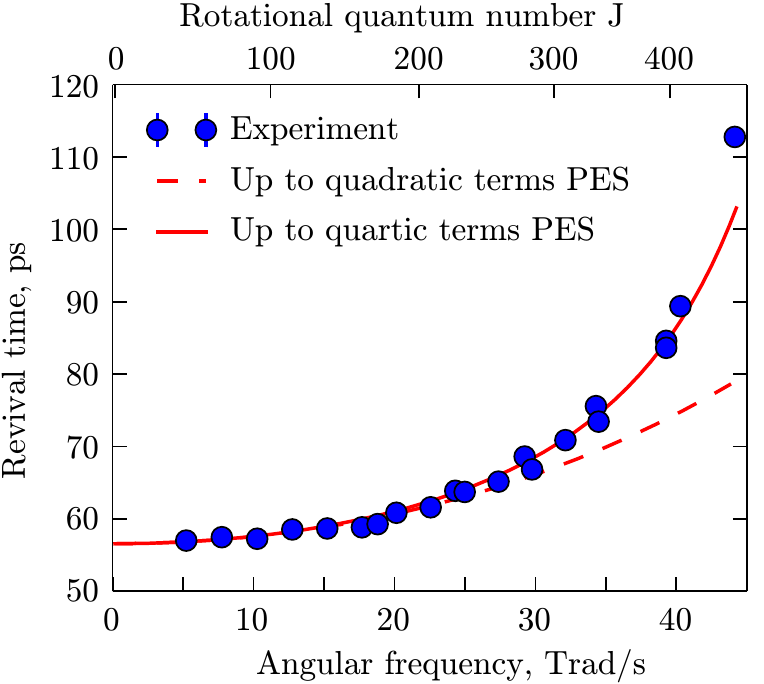}
    \caption{(color online) Revival period as a function of the rotational frequency of \sot{}, with the rotational quantum numbers shown along the upper horizontal axis. Experimental data (blue circles) are compared with the results of classical calculations, in which the potential energy surface (PES) is expanded to second and fourth order in deformation coordinates (dashed and solid lines, respectively).}
    \label{fig_stretch}
\end{figure}

At higher angular frequencies, the rising centrifugal forces stretch and bend the molecule, distorting the two S-O bonds and the angle between them. This causes the moment of inertia, and hence the revival period, to increase, as evident from the experimental data in \figrefs{fig_stretch} (blue circles). To describe the effect of the centrifugal distortion on the revival period, we carried out the following classical calculations. For a given rotational state $J$, we minimized the total energy:
$$E(J, r_1, r_2, \alpha)=\frac{\hbar^2J^2}{2I_c(r_1,r_2,\alpha)}+V(r_1, r_2, \alpha),$$
over the S-O bonds lengths $r_{1,2}$ and O-S-O angle $\alpha$. The potential energy surface (PES) $V$ of \sot{} was expanded in Taylor series up to the forth order in deformations $\delta x_i=r_1-r_e, r_2-r_e, \alpha-\alpha_e$, where $r_e$ and $\alpha_e$ are the values of $r_{1,2}$ and $\alpha$ for the molecule at rest:
\begin{multline*}
V(r_1, r_2, \alpha)=V(r_e, r_e, \alpha_e)+\frac{1}{2}\sum_{ij}f_{ij}\delta x_i\delta x_j+\\
\frac{1}{6}\sum_{ijk}f_{ijk}\delta x_i\delta x_j\delta x_k
+\frac{1}{24}\sum_{ijkl}f_{ijkl}\delta x_i\delta x_j\delta x_k\delta x_l,
\end{multline*}
with the corresponding force constants $f$ taken from Ref.\citenum{Martin1998}. We then determined the classical angular velocity and the revival period for a given $J$ as $$\omega(J)=\left[ E(J+1)-E(J) \right]/\hbar$$
and
$$T_\text{rev}=2\pi \hbar \left[E(J-1)-2E(J)+E(J+1)\right]^{-1},$$
respectively. The results of these calculations are shown in \figrefs{fig_stretch}. Expanding PES to second order in deformations $\delta x_i$ failed to explain the experimental observations at frequencies higher than $3\times10^{13}$~rad/s (dashed line). The quartic expansion, on the other hand, proved sufficient (solid line). At the highest achieved rotational frequencies of $4.4\times10^{13}$~rad/s, the calculated bending angle and bond stretching reached $10\degree$ and $10~\mathrm{nm}$, respectively. Both bonds were found to stretch by an equal amount, as could be anticipated from the symmetry of the system.

In summary, we demonstrated and studied a new mechanism of field-free alignment of asymmetric top molecules with an optical centrifuge. An intuitive classical picture has been presented to explain the observed behavior in the gas of sulfur dioxide. Owing to the adiabaticity of the centrifuge spinning, it leaves \sot{} rotating with its O-S-O plane aligned with the plane of rotation long after the excitation pulse is gone. This planar alignment reduces complex rotational dynamics of an asymmetric top to that of a simple linear rotor. Similarly to the latter, centrifuged \sot{} molecules exhibit periodic revivals of the transient field-free three-dimensional alignment. The observed FF3DA stems from the slight deviation of the centrifuge action from fully adiabatic. To enhance the degree of the three-dimensional alignment, one could add a short non-adiabatic rotational kick in the plane of the centrifuge-induced rotation. By measuring the revival period as a function of the rotational frequency, we examined the centrifugal distortion of \sot{}. At extreme levels of rotational excitation, available with an optical centrifuge, the molecule bends by up to ten degrees. Both the planar alignment and the tuneable centrifugal bending add to the arsenal of laser-based tools for controlling molecular dynamics.

This research has been supported by the grants from CFI, BCKDF and NSERC and carried out under the auspices of the UBC Center for Research on Ultra-Cold Systems (CRUCS).

\end{document}